\documentclass[a4paper,11pt]{article}
\usepackage{pos}

\renewcommand{\hookAfterAbstract}{%
\par\bigskip\bigskip
\textsc{IPPP/23/50}
}

\usepackage{tikz}
\usepackage{wrapfig}
\newcommand\mcmule{\textsc{McMule}}
\newcommand\M[2]{\mathcal{M}_{#1}^{(#2)}}
\newcommand\fM[2]{\mathcal{M}_{#1}^{(#2)f}}

\def\D{{\rm d}}
\def\eik{\mathcal{E}}
\def\ieik{\hat{\mathcal{E}}}

\title{QED at NNLO and beyond for precision experiments}

\author*[a]{Yannick Ulrich}

\affiliation[a]{Institute for Particle Physics Phenomenology, Department of Physics, \\
Durham University, Durham, DH1 3LE, UK}

\emailAdd{yannick.ulrich@durham.ac.uk}

\abstract{
    Low-energy experiments allow for some of the most precise measurements in particle physics, such as $g-2$.
    To make the most of these experiments, theory needs to match the experimental precision.
    Over the last decade, this meant that even in QED next-to-next-to-leading order calculations (or even more in some cases) became necessary.
    McMule (Monte Carlo for MUons and other LEptons) is a framework that we have developed to obtain NNLO predictions for a number of processes, such as $e\mu\to e\mu$, $ee\to ee$, and $\mu\to e\nu\bar\nu$.
    I will discuss some of the challenges faced when dealing with QED corrections and some possible solutions we have implemented in McMule, namely the subtraction scheme FKS$^\ell$, massification, and next-to-soft stabilisation.
    I will also demonstrate how to calculate the three-loop massification constant that will be required at N$^3$LO.
}

\FullConference{%
    16th International Symposium on Radiative Corrections: Applications of Quantum Field Theory to Phenomenology (RADCOR2023)\\
    28th May - 2nd June, 2023\\
    Crieff, Scotland, UK
}

\begin{document}
\maketitle

\section{Introduction}

Current and future low-energy precision experiments such as MUonE~\cite{Abbiendi:2016xup} ($e\mu\to e\mu$, PRad~\cite{Xiong:2019umf,PRad:2020oor} and MUSE~\cite{MUSE:2017dod,Cline:2021ehf} ($\ell p\to \ell p$) will require unprecedented precision from the theory side to reach their full potential.
The dominant corrections in these experiments come from QED rather than QCD, which only enters non-perturbatively in the description of protons and pions.
The level of precision reached by these experiments necessitates going beyond next-to-leading order (NLO) to at least next-to-next-to-leading order (NNLO).

\mcmule{} is a framework to perform these calculations.
In Table~\ref{tab:processes}, we list the processes currently implemented in \mcmule{} as well the order at which they are known.
In these proceedings, we will briefly review the tools used in \mcmule{} to make these calculations possible and efficient.
These include the subtraction scheme (Section~\ref{sec:fks}), massification (Section~\ref{sec:massification}), and next-to-soft (NTS) stabilisation (Section~\ref{sec:nts}).
Next, we will discuss steps towards N$^3$LO in Section~\ref{sec:n3lo} before concluding in Section~\ref{sec:conclusion}.

\begin{table}[b]
    \centering
    \footnotesize
    \begin{tabular}{l|p{2.2cm}|l|l}
        \bf Process                 & \bf order          & \bf comment & \bf reference \\\hline
        $\mu\to e\nu\bar\nu$        & NNLO               & polarised   & \cite{Engel:2019nfw} \\
        $\mu\to e\nu\bar\nu \gamma$ & NLO                & polarised   & \cite{Pruna:2017upz} \\
        $\mu\to e\nu\bar\nu ee$     & NLO                & polarised   & \cite{Pruna:2016spf} \\
        $\mu\to e X$                & NLO                & polarised   & \cite{Banerjee:2022nbr}\\
        \hline
        $\ell p\to\ell p$           & NNLO               & FFs at NLO  & \cite{Engel:2023arz} \\
        $e\mu \to e\mu$             & NNLO               & massified   & \cite{Broggio:2022htr} \\
        $ee\to\tau\tau$             & dom. NNLO + NLO EW & polarised   & \cite{Kollatzsch:2022bqa} \\
        \hline
        $e^+e^-\to e^+e^-$          & ph. NNLO           & massified   & \cite{Banerjee:2021mty} \\
        $e^-e^-\to e^-e^-$          & NNLO               & massified   & \cite{Banerjee:2021qvi} \\
        \hline
        $ee\to\gamma\gamma$         & ph. NNLO           & massified   & \cite{Naterop:2021msc} \\
    \end{tabular}
    \caption{
        Processes implemented in \mcmule{}.
        ``dom.'' implies that only the dominant corrections are implemented (see eg.~\cite{Banerjee:2020tdt} for a precise definition).
        ``ph.'' means only photonic corrections (i.e. those without closed fermion loops) are implemented.
        $\ell p\to\ell p$ is implemented with form factors at NLO as indicated in the comments columns.
    }
    \label{tab:processes}
\end{table}

\section{\texorpdfstring{FKS$^\ell$}{FKSl} subtraction}
\label{sec:fks}

As part of the calculation of higher-order corrections, we naturally have to include divergent real corrections.
Since we want to perform this integration numerically, we need a prescription to deal with these singularities.
In QED, one typically treats fermions as massive rather than massless which is common in QCD calculations.
This means that we only have to treat soft singularities, i.e. $E_\gamma = \sqrt{s}/2 \times \xi \to 0$, with the collinear ones being regulated by the fermion mass.

The simple structure of soft singularities was demonstrated by Yennie, Frautschi, and Suura (YFS) in their seminal paper~\cite{Yennie:1961ad}.
When considering a process with a real photon emission, the matrix element (squared) can be approximated for soft photons
\begin{align}
    \M{n+1}\ell = \eik\,\M{n}\ell + \mathcal{O}(\xi)\,.
\end{align}
Here, we use $\M n\ell$ to denote the $\ell$-loop $n$-particle matrix element (squared) and defined the rescaled photon energy $\xi=2E_\gamma/\sqrt s$.
We have further defined the eikonal factor $\eik$ that encodes the angular distribution of the emitted photon which can be constructed trivially as
\begin{align}
    \eik = -\sum_{ij} Q_iQ_j\frac{p_i\cdot p_j}{(p_i\cdot p_\gamma)(p_j\cdot p_\gamma)}\,,
\end{align}
with fermion momenta (charges) $p_i$ ($Q_i$).
For simplicity, we have assumed all fermion momenta are incoming; the realistic case can be obtained by setting $p_i\to -p_i$ as required.
Integrating $\eik$ over the one-particle phase space, we get the integrated eikonal $\ieik$ which contains the (dimensionally regulated) soft singularity.
Naturally, the singularity in the $\ieik$ will exactly match the one in the one-loop virtual contribution $\M n1$.
However, the statement of YFS is even stronger.
All virtual singularities are subtracted to all loop orders by $\ieik$
\begin{align}
    e^{\ieik} \sum_{\ell=0}^\infty \M n\ell = \sum_{\ell=0}^\infty \fM n\ell\,,
\end{align}
where we have introduced the finite, eikonal-subtracted matrix element $\fM n\ell$.

We make use of this to construct the FKS$^\ell$ scheme~\cite{Engel:2019nfw} that is based on the original FKS scheme~\cite{Frixione:1995ms,Frederix:2009yq}.
The resulting subtraction scheme works at all orders in perturbation theory.
We begin by reviewing the basics of the FKS scheme at NLO before discussing the all-order statement.
For details of the derivation, we refer the reader to~\cite{Engel:2019nfw}.

At NLO, we have two contributions
\begin{align}
    \sigma^{(1)}
        = \int \D\Phi_n\ \M n1
        + \int \D\Phi_{n+1}\ \M {n+1}0\,.
\end{align}
Both the flux factor and the measurement function are implicit for simplicity.
In the type of slicing scheme that is commonly used in QED calculations, one introduces a cut-off $\xi_s=2E_s/\sqrt{s}$ to split the second integral
\begin{align}
    \sigma^{(1)}
        = \int \D\Phi_n\ \M n1
        + \int_0^{\xi_s} \D\Phi_{n+1}\ \M {n+1}0
        + \int_{\xi_s} \D\Phi_{n+1}\ \M {n+1}0\,.
\end{align}
In the second integral, $\M {n+1}0$ is approximated as $\eik\M n0$ so that the integrated eikonal can be used.
This results in
\begin{align}
    \sigma^{(1)}
        = \int \D\Phi_n\ \underbrace{\Big(\M n1 + \ieik \M n0\Big)}_{\fM n1}
        + \int_{\xi_s} \D\Phi_{n+1}\ \M {n+1}0\,,
\end{align}
which is manifestly finite.
However, since $\xi_s$ needs to be chosen small enough for the eikonal approximation to be valid, the numerical integration over $\D\Phi_{n+1}$ can be challenging.

In FKS, we instead subtract and add back a counter term that is constructed from $\eik$ as follows
\begin{align}
    \sigma^{(1)}
        = \int \D\Phi_n\ \underbrace{\Big(\M n1 + \ieik(\xi_c) \M n0\Big)}_{\fM n1}
        + \int \D\Phi_{n+1}\ \Big(\frac1\xi\Big)_c (\xi \M{n+1}0) \,,
    \label{eq:fks:nlo}
\end{align}
where we have defined the $c$-distribution acting on $\xi \M{n+1}0$
\begin{align}
    \int_0^1\D\xi\, \Big(\frac1\xi\Big)_c\, f(\xi)
    \equiv
    \int_0^1\D\xi\,\frac{f(\xi)-f(0)\theta(\xi_c-\xi)}\xi\,.
\end{align}
Here, we have introduced an unphysical parameter $\xi_c$ that can easily be confused with the slicing parameter $\xi_s$.
However, $\xi_c$ does not need to be small since it merely controls when subtraction takes place.
Varying it serves as a useful check for the implementation, but it can be chosen $\xi\sim0.3$ which alleviates the numerical issues of the slicing approach.
The prescription in \eqref{eq:fks:nlo} is finite as the distribution regulates the singular behaviour of $\M{n+1}0$ in the limit $\xi\to0$.

We can extend this scheme fairly easily to all-orders (cf.~\cite{Engel:2019nfw} for the explicit construction at NNLO and N$^3$LO)
\begin{align}
    \sigma^{(\ell)} &= \sum_{j=0}^\ell\int\D\Phi_{n+j}\frac1{j!}
        \Bigg[\prod_{i=1}^j \Big(\frac1{\xi_i}\Big)_c \Bigg]
        \fM{n+1}{\ell-j}
        \,,\\
    \fM m\ell &= \sum_{j=0}^\ell\frac{\ieik^j}{j!} \M m{\ell-j}\,.
\end{align}
This scheme has been successfully used for all NNLO calculations in \mcmule{} (cf. Table~\ref{tab:processes}).
We will come back to N$^3$LO application in Section~\ref{sec:n3lo}.

\section{Massification}
\label{sec:massification}

In the previous section, we have exploited the fact that fermions are massive to simplify the infrared structure.
However, this comes at a not insignificant cost when computing the required matrix elements.
While much progress has been made, including methods presented at this very conference, the calculation of the two-loop $\M n2$ with full mass dependence is still extremely challenging.
Luckily, it is rarely needed since the electron mass $m_e$ is much smaller than most other masses (such as the muon or proton masses $m_\mu$ or $m_p$, respectively) or kinematic variables (such as $s$, $t$, and $u$), collectively denoted as $S$.
This means that a leading power (LP) approximation in $m_e^2/S^2$ is often more than sufficient.
The error introduced by dropping these polynomially suppressed terms at two-loop is usually $\mathcal{O}(10^{-3})$ on the NNLO coefficient or $\mathcal{O}(10^{-5})$ on the total cross section.

Massification is a SCET-inspired factorisation of the amplitude into hard, collinear, and soft modes
\begin{align}
    \mathcal{A}(m_e) = \mathcal{S} \times \prod_{i}\sqrt{Z_i} \times \mathcal{A}(0) + \mathcal{O}(m_e)\,.
    \label{eq:massification}
\end{align}
The hard contribution corresponds to the amplitude $\mathcal{A}(0)$ where $m_e=0$ and can usually be obtained from existing calculations.
The collinear modes are accounted for by a factor of $\sqrt{Z_i}$ for each light external particle.
They are not process dependent and can be re-used between different calculations after obtaining them in a matching calculation.
Finally, the soft contribution $\mathcal{S}$ is process dependent, but one can show that only diagrams with closed electron loops contribute.

This means that, once $Z_i$ is known, we only need to consider closed fermion loops for $\mathcal{S}$ which significantly reduces the complexity of obtaining (and eventually evaluating) $\mathcal{A}(m_e)$.
However, since \mcmule{} deals with energy scales below 10 GeV, we also need to consider hadronic loops.
These cannot be calculated perturbatively and instead need to be obtained from data and integrated numerically.
\mcmule{} uses either a dispersive approach~\cite{Jegerlehner:2017gek} or a hyperspherical one~\cite{Levine:1974xh,Levine:1975jz} for these integrals.
Since a numerical integration is required anyway, we also use it to calculate closed electron loops with full dependence on all masses.
This means that we never have to analytically calculate $\mathcal{S}$.

As mentioned before, the $\sqrt{Z_i}$ are universal and can be calculated in the method of regions as was done in~\cite{Engel:2018fsb}.
However, a simpler approach is to simply solve~\eqref{eq:massification} for $Z_i$ for a simple process where both $\mathcal{A}(m_e)$ and $\mathcal{A}(0)$ are known.
The form factors of $\gamma^*(Q^2)\to ee$ are the ideal environment for this as they were recently calculated semi-numerically to three-loop.
By expanding the form factors for $Q^2\gg m_e^2$ we can obtain the photonic parts of $Z$ at three-loop accuracy
\begin{align}
    Z\Big|_{\rm ph.} = \frac{\mathcal{A}(Q^2\gg m_e^2) }{\mathcal{A}(0)}\Big|_{\rm ph.}\,.
    \label{eq:Zsolved}
\end{align}
Using the result of~\cite{Gehrmann:2010ue} for $\mathcal{A}(0)$ and of~\cite{Ablinger:2017hst,Fael:2022rgm} for $\mathcal{A}(Q^2\gg m_e^2)$ we can obtain an analytic answer for $Z$ except for the finite three-loop part which contains a numeric constant which is known to high accuracy.
The full expression for $Z$ in the convention of \cite{Engel:2018fsb} can be found in Appendix~\ref{sec:Z} and attached to this submission in electronic form.
Even though~\eqref{eq:Zexpl} contains a numerical constant rather than just transcendental constants, it can still be used for calculations.
It will eventually allow us to perform approximate N$^3$LO calculations for processes where the full mass dependence of the triple-virtual is firmly out of reach such as $ee\to ee$.

\section{Next-to-soft stabilisation}
\label{sec:nts}

For the real-virtual corrections we need the real-virtual matrix element $\M{n+1}1$.
Thanks to the tremendous progress made in the automation of one-loop calculations, obtaining this is fairly straightforward using OpenLoops~\cite{Buccioni:2017yxi,Buccioni:2019sur} and Collier~\cite{Denner:2016kdg}.
However, the vast majority of phase space points that need to be evaluated will have a soft and/or collinear photon.
While OpenLoops can handle these, the numerical stability and speed suffer as the rescue and stability system is not well suited for QED calculations.

A way to solve this problem is to expand $\M{n+1}1$ in the problematic region, namely for soft emission $\xi\to0$.
At LP, this results in the well-known eikonal we have encountered in Section~\ref{sec:fks}.
Extending the expansion to next-to-leading power (NLP), i.e. also include terms that $\mathcal{O}(\xi^{-1})$, allows us to use the expansion earlier, improving speed and stability.

At tree-level, the universal nature of this next-to-soft (NTS) expansion was proven by Low, Burnett, and Kroll~\cite{Low:1958sn,Burnett:1967km} (LBK) for unpolarised scattering.
This was later extended first to one-loop~\cite{Engel:2021ccn} and then to all-orders~\cite{Engel:2023ifn}, though still only for QED with massive fermions
\begin{align}
    \begin{split}
        \mathcal{M}_{n+1} = \Bigg[\frac1{\xi^2}\eik + \frac1\xi
            \sum_{ij}Q_iQ_j \frac{p_i\cdot \tilde D_j}{p_i\cdot p_\gamma}
            &+\frac1\xi\sum_{i,j,k\neq j}Q_kQ_j^2Q_i\Big(
                    \frac{p_j\cdot p_i}{(p_\gamma\cdot p_j)(p_\gamma\cdot p_i)}
            \\&
                -\frac{p_k\cdot p_i}{(p_\gamma\cdot p_k)(p_\gamma\cdot p_i)}
            \Big)2S^{(1)}(p_j, p_k, p_\gamma)
            + \mathcal{O}(\xi^0)
            \Bigg]\mathcal{M}_n\,.
    \end{split}
    \label{eq:nts}
\end{align}
Here, we have introduced the LBK differential operator $\tilde D_j$
\begin{align}
    \tilde D_j^\mu = \sum_L\Big(
        \frac{p_j^\mu}{p_j\cdot p_\gamma}
        p_\gamma\cdot \frac{\partial s_L}{p_j}
        - \frac{\partial s_L}{\partial p_j^\mu}
    \Big) \frac{\partial}{\partial s_L}
\end{align}
that takes derivatives of the non-radiative matrix element w.r.t. invariants $s_L$ (see~\cite{Engel:2023ifn} for further information on this).
We have also introduced the one-loop exact soft function $S^{(1)}(p_j, p_k, p_\gamma)$ which can be found calculated in~\cite{Engel:2021ccn}.
\begin{figure}
    \centering
    \includegraphics[width=0.8\textwidth]{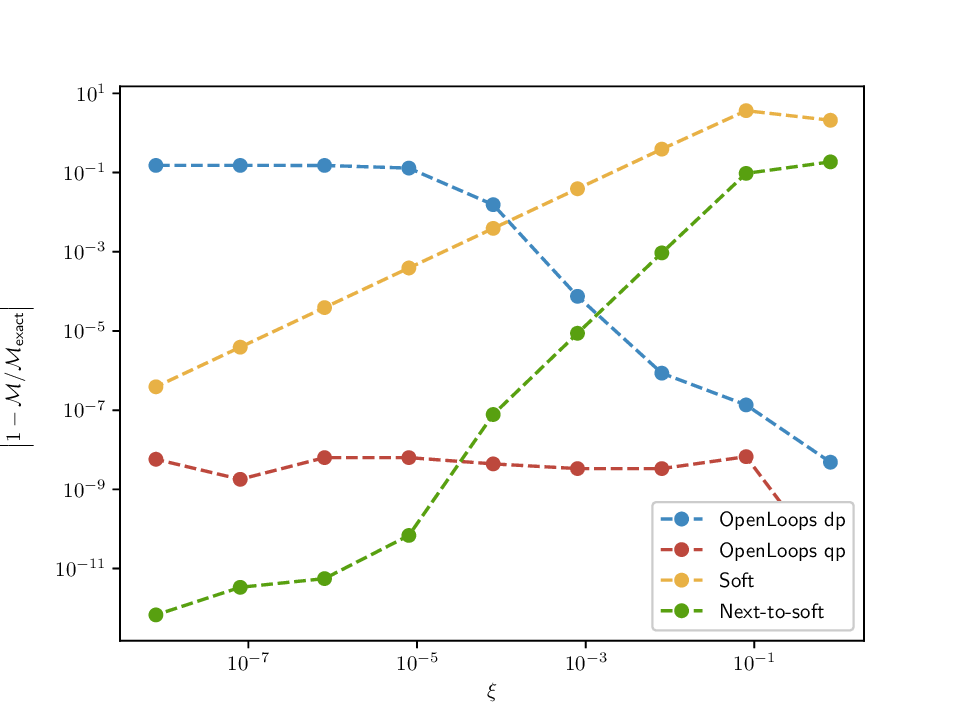}
    \caption{
        The different implementations of the one-loop matrix element for $ee\to ee\gamma$
    }
    \label{fig:nts}
\end{figure}

To study these approximations, we consider the process $e^+e^-\to e^+e^-\gamma$ at one-loop and compare an arbitrary precision calculation with OpenLoops in double precision, OpenLoops in quadruple precision, the LP expansion, and the NLP expansion.
The results are shown in Figure~\ref{fig:nts}.
One can clearly see that double precision is insufficient around $\xi=10^{-3}$ while quadruple precision is always acceptable.
However, the LP approximation only becomes reliable around $\xi=10^{-7}$.
The NLP expansion allows us to construct an implementation of the matrix element that is always stable and much faster than quadruple precision by using double precision OpenLoops up to $\xi\simeq 10^{-3}$ before switching to the NLP expansion of \eqref{eq:nts}.
This was first demonstrated in~\cite{Banerjee:2021mty,Banerjee:2021qvi} in the context of $ee\to ee$ where it allowed us to implement the first NNLO calculation of Bhabha and M\o{}ller scattering.
It was later used for $ee\to\tau\tau$~\cite{Kollatzsch:2022bqa} and $e\mu\to e\mu$~\cite{Broggio:2022htr}.

It is possible to extend~\eqref{eq:nts} to polarised processes at the cost of simplicity.
For real-virtual corrections, \eqref{eq:nts} only depends on the one-loop and tree-level reduced matrix elements $\M n1$ and $\M n0$.
It was shown in~\cite{Kollatzsch:2022bqa} that when considering this expansion in the method of regions, the hard region gets an additional contribution from the polarisation.

\section{Towards \texorpdfstring{N$^3$LO}{N3LO}}
\label{sec:n3lo}

While we have developed these tools for NNLO calculation, they can all be taken to N$^3$LO.
The subtraction scheme works at all orders, as does NTS stabilisation.
Of course, the main bottleneck is the availability of the matrix elements which makes true $2\to2$ processes at N$^3$LO extremely difficult in QED.
However, we can consider eg. only the initial-state corrections to $ee\to\gamma^*(\to \mu\mu, \pi\pi, ...)$ or similarly the electron-line corrections~\cite{Banerjee:2020tdt} to $\mu$-$e$ scattering.
This is possible because the heavy-quark form factor is known at three-loop accuracy which constitutes the triple-virtual corrections of these (sub)processes.

\begin{wrapfigure}{r}{0.31\textwidth}
    \centering
    \scalebox{0.8}{
        \begin{tikzpicture}
            \draw[->](0,0) -- (0,4) node[right] {$\theta$};
            \draw[->](0,0) -- (4,0) node[right] {$\xi$};
            \draw[dashed](1,0) -- (1,4);
            \draw[dashed](1,1) -- (4,1);
            \node at (0.5,2){NTS};
            \node at (2.5,0.5){jettification};
            \node at (2.5,2){massification};
        \end{tikzpicture}
    }
    \caption{
        The arrangement of the different approximations for $\M{n+1}2$.
    }
    \label{fig:rvv}
\end{wrapfigure}
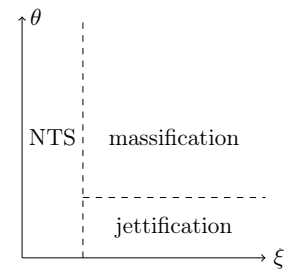
The real-virtual-virtual corrections, i.e. $ee\to\gamma\gamma^*$, have been known in the limit $m_e\to0$ for many years as part of the NNLO corrections to three-jet production~\cite{Gehrmann:2000zt,Gehrmann:2001ck} and have recently been recalculated~\cite{Badger:2023xtl,Fadin:2023phc}.
Massification allows us to recover the mass effects in the bulk of the phase space where $m_e^2 \ll S^2$.
However, for soft and/or collinear emission, $p_i\cdot p_\gamma$ may become comparable or even smaller than $m_e$.
For soft emission, the NLP expansion of~\eqref{eq:nts} is sufficient to address this issue.
For collinear emission, a similar expansion which we call jettification is needed.
This has been demonstrated at LP and one-loop in~\cite{Engel:2021ccn}.
Extending this to two-loop is the last missing ingredient for the real-virtual-virtual.
The interplay between the different expansion is shown in Figure~\ref{fig:rvv}.

With NTS stabilisation and OpenLoops, the real-real-virtual corrections should be feasible as well once NTS stabilisation is extended to two soft emissions.
The last remaining part of the calculation, the triple-real correction, is unlikely to form a bottleneck either.

\section{Conclusion}
\label{sec:conclusion}

We have reviewed the important theoretical underpinnings of the \mcmule{} framework.
Phenomenological results for $\mu$-$e$ scattering can be found elsewhere in these proceedings~\cite{Rocco:2023mule}.
These tools allow for the systematic calculation of NNLO corrections in QED, similar to what happened in QCD some years ago.
We can now say with confidence that the NNLO era has arrived, not only for QCD but also for QED with massive fermions.

\paragraph{Acknowledgement}
I acknowledge support by the UK Science and Technology Facilities Council (STFC) under grant ST/T001011/1.
I would like to thank Fabian Lange and Kay Sch\"o{}nwald for their help extracting the relevant parts of $\mathcal{A}(Q^2\gg m_e^2)$.
Finally, a huge thank you to my collogues in the \mcmule{} Team for their support developing and implementing this framework.

\appendix

\section{The massification constant at three-loop}
\label{sec:Z}

The equation~\eqref{eq:Zsolved} evaluates to
\def\lineA{\\&\qquad}
\def\lineB{\\&\qquad\qquad}
\begin{align}\begin{split}
    \sqrt{Z_i} = 1
    &+ a \Bigg[
        \frac{1}{\epsilon^2}
        + \frac{1}{2\epsilon}
        + \zeta_2 + 2
        + \Big(4 + \frac{1}{2} \zeta_2\Big) \epsilon
        + \Big(8 + 2 \zeta_2 + \frac{7}{4} \zeta_4\Big) \epsilon^2
        \lineA
        + \Big(16 + 4 \zeta_2 + \frac78 \zeta_4\Big) \epsilon^3
        + (32 + 8 \zeta_2 + \frac{7}{2} \zeta_4 + \frac{31}{16} \zeta_6\Big) \epsilon^4
        +\mathcal{O}(\epsilon^5)
    \Bigg]
    \\&
    + a^2 \Bigg[
        \frac{1}{2\epsilon^4}
        + \frac{1}{2\epsilon^3}
        + \frac{1}{\epsilon^2} \Big(\frac{51}{24} + \zeta_2\Big)
        + \frac{1}{\epsilon}\Big(\frac{43}{8} - 2 \zeta_2 + 6 \zeta_3\Big)
        \lineA
        + \frac{369}{16} + \frac{61}{4} \zeta_2 - 18 \zeta_4 - 24 \zeta_2\ \log2 - 3 \zeta_3
        \lineA
        + \Big(
            -\frac{173}{32}
            + \frac{221}{4} \zeta_2
            - 12 \zeta_2\ \log2 + 49 \zeta_3
            + 4 \log^4 2 + 48 \log^2 2\ \zeta_2 + 96 a_4 \lineB - \frac{351}{2} \zeta_4
            - 18 \zeta_2\ \zeta_3 - 3 \zeta_5
            \Big)\epsilon
        \lineA
        + \Big(
            \frac{2841}{64}
            + \frac{2751}{8} \zeta_2
            - 288 \zeta_2\ \log2 + 161 \zeta_3
            + 2 \log^4 2 + 24 \zeta_2\ \log^2 2 + 48 a_4 \lineB + \frac{387}{4} \zeta_4
            - \frac{24}{5} \log^5 2 + 252 \zeta_4\ \log2 - 87 \zeta_2\ \zeta_3 + 576 a_5 - \frac{1431}{2} \zeta_5
            \lineB
            + 4 \zeta_2\ \log^4 2 - 96 \zeta_3\ \log^3 2 - 60 \zeta_4\ \log^2 2 + 84 \zeta_2\ \zeta_3\ \log2 + 96 \zeta_2\ a_4 \lineB - \frac{81}{2} \zeta_3^2 - 613 \zeta_6
            \Big) \epsilon^2
            +\mathcal{O}(\epsilon^3)
    \Bigg] \\&
    + a^3 \Bigg[
        \frac{1}{6\epsilon^6}
        + \frac{1}{4\epsilon^5}
        + \frac{1}{\epsilon^4} \Big(\frac{9}{8} + \frac{1}{2}\zeta_2\Big)
        + \frac{1}{\epsilon^3} \Big(\frac{163}{48} - \frac{9}{4} \zeta_2 + 6 \zeta_3\Big)
        \lineA
        + \frac{1}{\epsilon^2} \Big(\frac{39}{2} + \frac{103}{8} \zeta_2 - 24  \zeta_2\ \log2 - \frac{151}{8} \zeta_4\Big)
        \lineA
        + \frac{1}{\epsilon} \Big(
            -\frac{77}{24}
            + \frac{915}{16} \zeta_2
            - 24 \zeta_2\ \log2 + \frac{425}{6} \zeta_3
            + 4 \log^4 2 + 48 \zeta_2\ \log^2 2 + 96 a_4 \lineB - \frac{2709}{16} \zeta_4
            - \frac{52}{3} \zeta_2\ \zeta_3 - 43 \zeta_5
        \Big)
        \lineA
        - 342.0591735940860642547644580773479603199
        + \mathcal{O}(\epsilon)
    \Bigg] + \mathcal{O}(a^4)\,.
    \label{eq:Zexpl}
\end{split}\end{align}
We have used the conventional zeta function as well as $a_n = {\rm Li}_n(\tfrac12)$.
This result is also available in electronic form attached to this submission.
The numerical constant is exact to forty digits.

\bibliographystyle{JHEP}
\bibliography{mcmule-theory}

\end{document}